\documentclass[twocolumn,showpacs,preprintnumbers,amsmath,amssymb]{revtex4}
\usepackage[pdftex]{graphicx}

\begin{document}

\title{The electronic structure of {\em R}NiC$_2$ intermetallic compounds}

\author{J.~Laverock, T.~D.~Haynes, C.~Utfeld and S.~B.~Dugdale}

\affiliation{H.~H.~Wills Physics Laboratory, University of Bristol, Tyndall
Avenue, Bristol BS8 1TL, United Kingdom}

\begin{abstract}
First-principles calculations of the electronic structure of members of
the $R$NiC$_2$ series are presented, and their Fermi surfaces investigated
for nesting propensities which might be linked to the charge-density waves
exhibited by certain members of the series ($R$ = Sm, Gd and Nd). Calculations
of the generalized susceptibility, $\chi_{0}({\bf q},\omega)$, show strong
peaks at the same ${\bf q}$-vector in both the real and imaginary parts
for these compounds. Moreover, this peak occurs at a wavevector which is
very close to that experimentally observed in SmNiC$_2$.  In contrast, for
LaNiC$_2$ (which is a superconductor below 2.7K) as well as for ferromagnetic
SmNiC$_2$, there is no such sharp peak. This could explain the absence of
a charge-density wave transition in the former, and the destruction of the
charge-density wave that has been observed to accompany the onset of ferromagnetic
order in the latter.
\end{abstract}


\maketitle 

\section{Introduction}
The idea that the Fermi surface (FS), through instabilities in the electronic
structure (such as nesting features or van Hove saddle points), can drive
low-dimensional systems into new ground states such as spin-density or
charge-density waves (CDW), is well-known (for example, the transition
metal dichalcogenides \cite{wilson74} and Cr \cite{overhauser62}).
More recent theoretical studies of the role of the FS in CDW formation
\cite{johannes06,johannes08} have contributed greatly to our understanding of
the phenomenon, and have clarified a number of misconceptions. Specifically,
those authors emphasize that simply inspecting the FS for possible
nesting features without actually calculating the real part of the
electronic susceptibility does not help in predicting CDW instabilities
\cite{johannes08}. They further urge caution about attributing such nesting as
the only (or even main) driving force for such instabilities. In this paper
we calculate the electronic structure, FS and susceptibilities
of several members of the $R$NiC$_2$ family which have been shown to
exhibit a fascinating array of electronic instabilities \cite{murase04,schafer97}.
SmNiC$_2$ has recently been found to host an interesting interplay between
CDW and ferromagnetic order \cite{shimomura09}. On cooling below a
temperature of 148K, a resistivity anomaly and the appearance of satellite
peaks in x-ray scattering indicate the formation of a CDW.  The critical
phonon softening, inferred from the thermal diffuse scattering above 148K
(and which also disappears at the ferromagnetic transition) occurs at two
specific wavevectors, namely ${\bf q}_{1} = (0.5, 0.52, 0)$ and ${\bf q}_{R}
= (0.5, 0.5, 0.5)$ \cite{shimomura09}.  Of these two wavevectors, it is
the incommensurate ${\bf q}_{1}$ modulation which develops into a CDW.
Below 17.7K the satellite peaks suddenly disappear and there is a sharp
decrease in the resistivity, and both phenomena are coincident with the
appearance of ferromagnetic order \cite{shimomura09}.  Shimomura {\it et al.}
tentatitively suggest that the diffuse scattering associated with the phonon
softening, the formation of the CDW state and their disappearance at the
(first order) ferromagnetic transition could be collectively understood from
a knowledge of the FS of SmNiC$_{2}$ \cite{shimomura09}.  In this paper we
address this issue through {\it ab initio} calculations of the electronic
structure of the $R$NiC$_2$ system.

\section{Electronic structure calculations}

\begin{figure}[b]
\includegraphics[width=1.00\linewidth,clip]{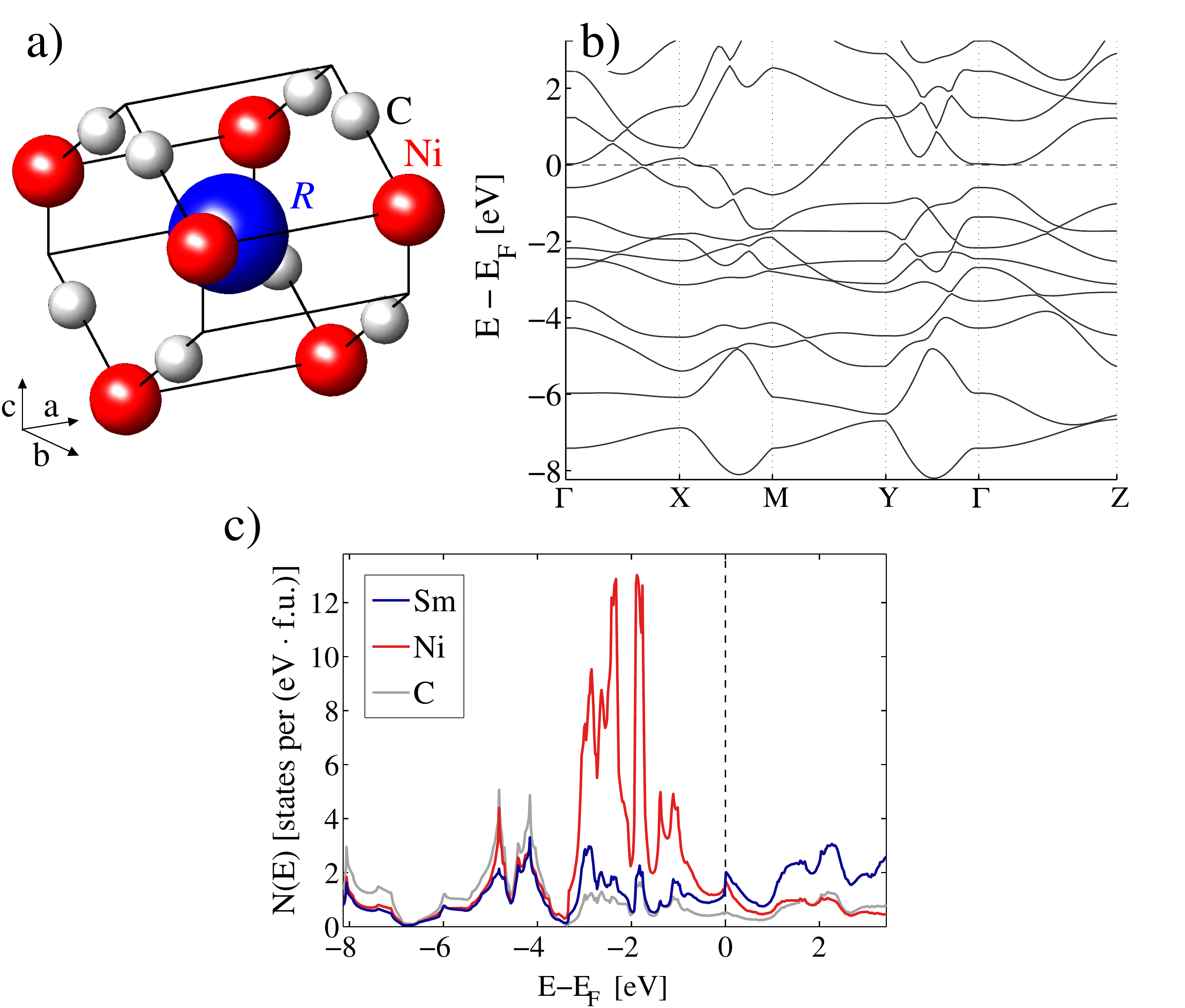}
\caption{\label{f:struc} (color online) The (a) crystal structure, (b)
electronic band structure, and (c) density of states of SmNiC$_2$.}
\end{figure}

\begin{table}[t!]
\begin{tabular}{|c|c|c|c|c|c|}
\hline
$R$       & a (\AA)  & b (\AA) & c (\AA) & Ref. & $N(E_{\rm F})$ \\
 & & & & & [states per (eV $\cdot$ f.u.)] \\
\hline
La & 3.956 & 4.563 & 6.204 & \cite{jeitschko86} & 4.34 \\
Nd & 3.765 & 4.544 & 6.146 & \cite{jeitschko86} & 2.64 \\
Sm & 3.703 & 4.529 & 6.098 & \cite{onodera98}   & 3.84 \\
Gd & 3.649 & 4.518 & 6.077 & \cite{jeitschko86} & 4.28 \\
\hline
\end{tabular}
\caption{Lattice parameters used in the electronic structure calculations,
accompanied by the references from which they were taken. The density of states
at the Fermi level, $N(E_{\rm F})$, obtained from the calculations, is also
given.}
\label{rareearthtable}
\end{table}

\begin{figure}[tb]
\includegraphics[width=1.00\linewidth,clip]{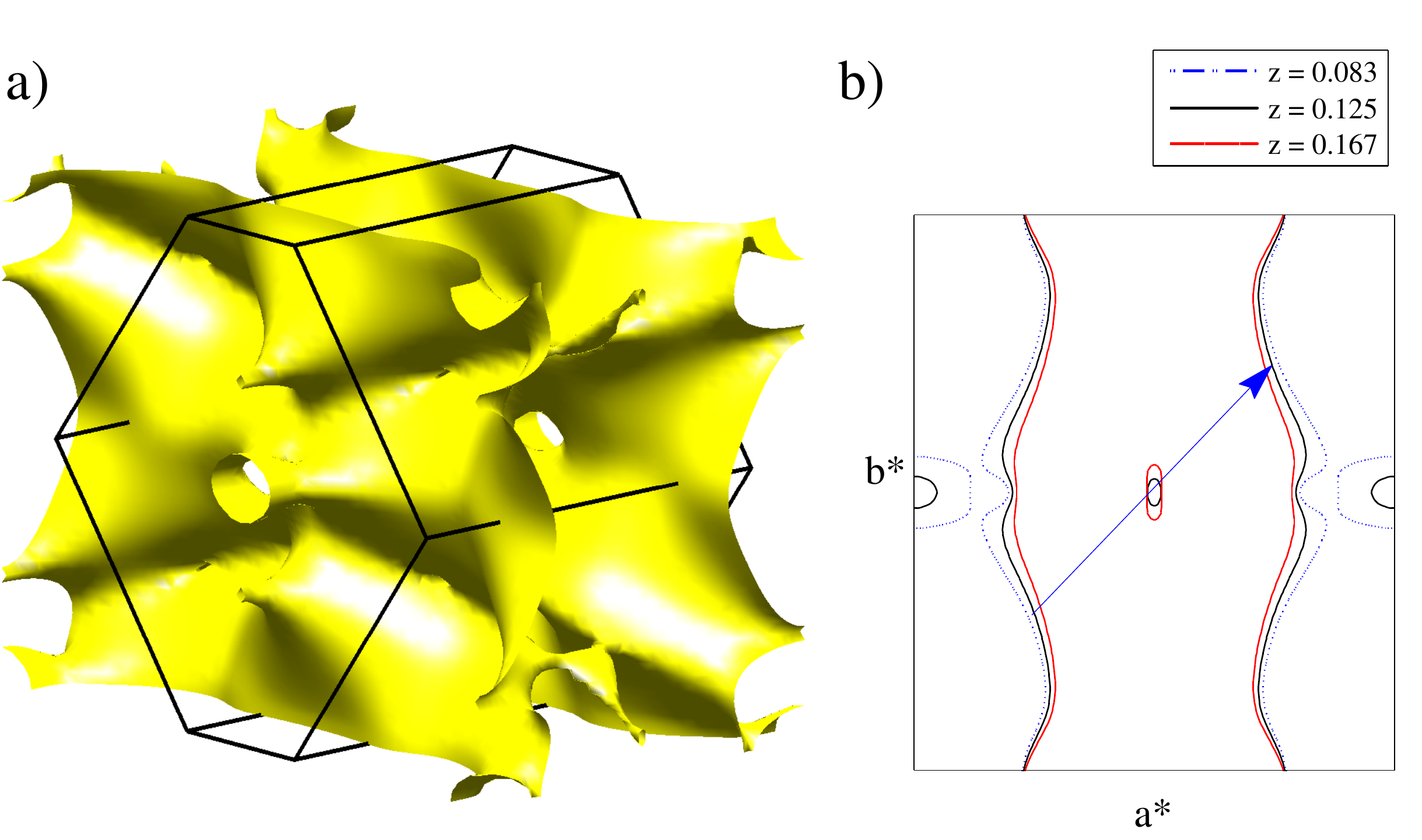}
\caption{\label{f:smfs} (color online) The FS of SmNiC$_2$, (a) in 3D with the
first Brillouin zone marked, and (b) several contours in planes perpendicular
to $c^*$. The arrow in (b) denotes the experimentally observed CDW wavevector
${\bf q} = (0.5, 0.52, 0)$. All units are in terms of $2\pi/(a,b,c)$.}
\end{figure}

The $R$NiC$_2$ family possess the orthorhombic CeNiC$_2$ structure
(Space Group 38, {\it Amm2})\cite{atpos}, which is shown in Fig.~\ref{f:struc}a. The
electronic band structure of $R$NiC$_2$ was calculated for several $R$ (La,
Nd, Sm and Gd) using the scalar-relativistic linear muffin-tin orbital (LMTO)
method within the atomic sphere approximation (ASA) including combined-correction
terms \cite{barbiellini03}. Here, the potential is described within the
local density approximation for non-magnetic calculations \cite{hedin71}
and the local spin density approximation for spin-polarized calculations
\cite{gunnarsson76}.  The lattice constants used in the calculations are shown
in Table~\ref{rareearthtable}.  All calculations included a basis of $s$, $p$,
$d$ and (for Ni and $R$) $f$ states; self-consistency was achieved at 1099
k-points in the irreducible $(1/8)^{\rm th}$ wedge of the BZ (corresponding
to a mesh of $ 13 \times 13 \times 13$ in the full BZ). The localized rare-earth
$f$-electrons were described using the open-core method, in which they are
treated as partially-filled core states.

\begin{figure}[tb]
\includegraphics[width=1.00\linewidth,clip]{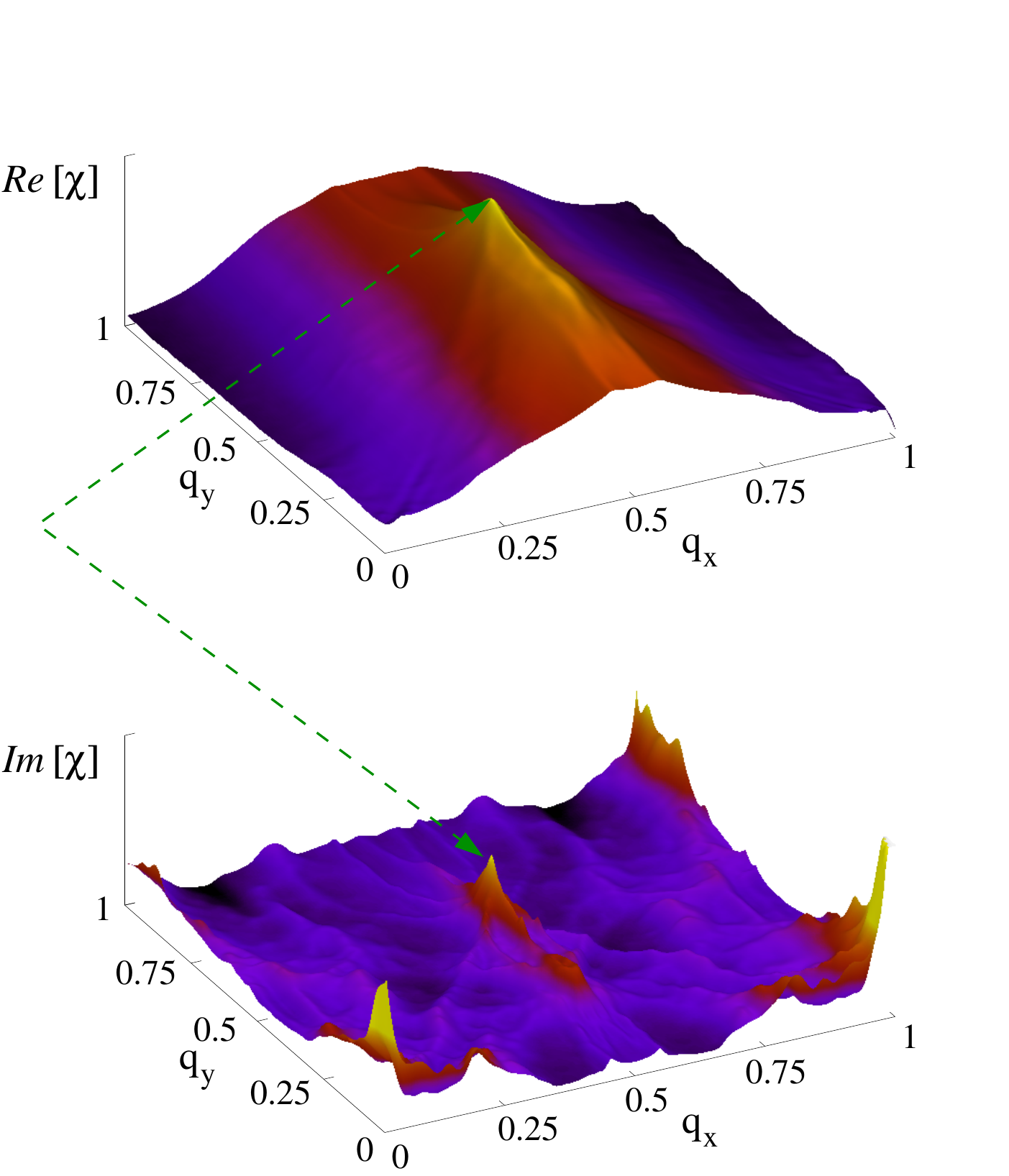}
\caption{\label{f:chi1} (color online) $\chi_0({\bf q})$ for SmNiC$_2$. The
arrow denotes the peak in both real and imaginary parts of $\chi_0({\bf q})$
at ${\bf q}~=~(0.5,~0.56,~0)$.}
\end{figure}

\begin{figure}[tb]
\includegraphics[width=1.0\linewidth,clip]{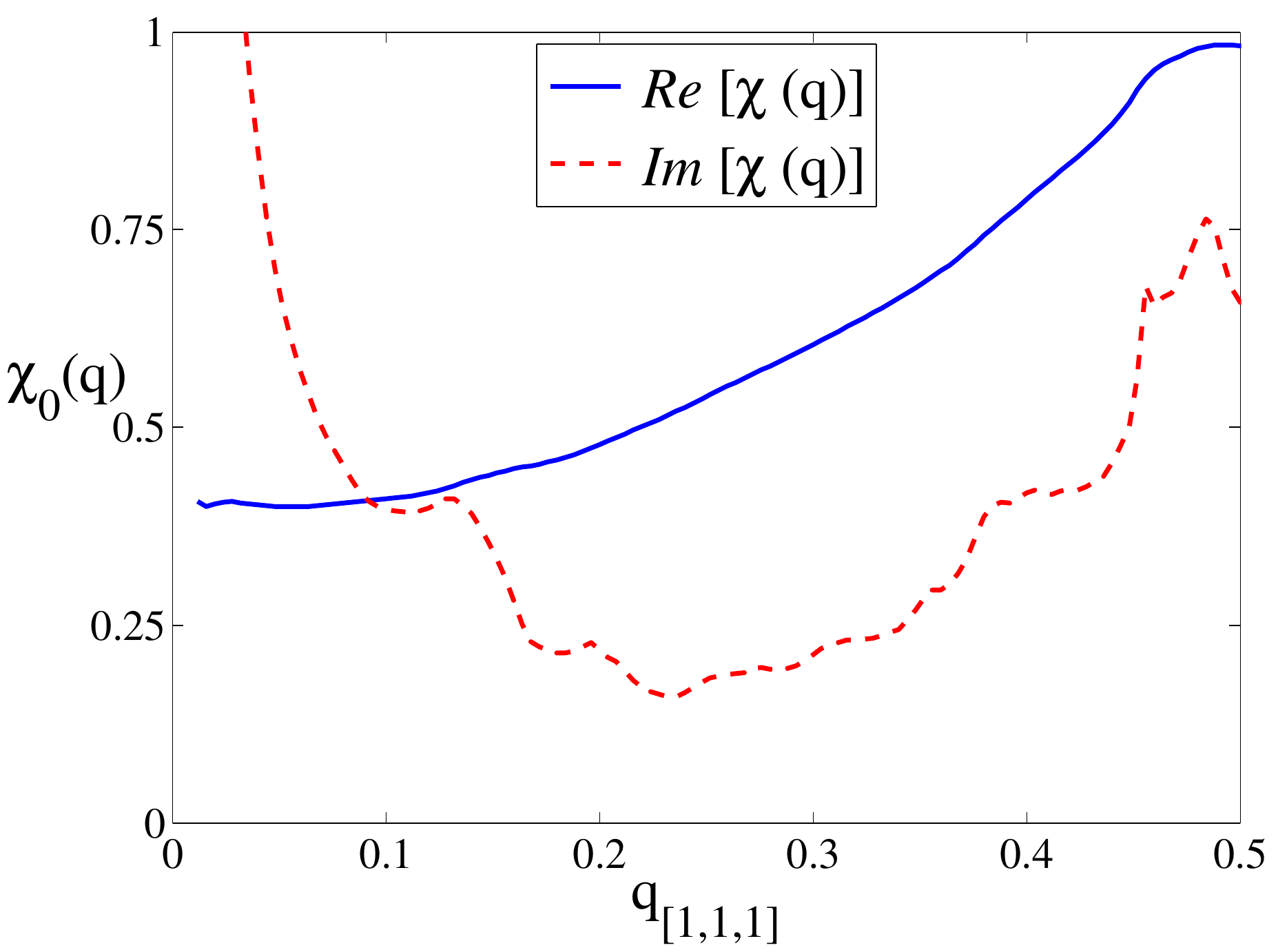}
\caption{\label{f:chi111} (color online) Real and imaginary parts of
$\chi_0({\bf q})$ along the $[1,1,1]$ direction, showing the peak at ${\bf
q}_{\rm R} = (0.5, 0.5, 0.5)$ associated with the thermal diffuse scattering
observed by Ref.\ \cite{shimomura09}. Owing to the crystal structure, there is a
symmetry axis at $q_{[1,1,1]} = (0.5, 0.5, 0.5)$.}
\end{figure}

The electronic band structure and density of states of SmNiC$_2$ are shown
in Fig.~\ref{f:struc}b and Fig.~\ref{f:struc}c, respectively. The FS,
shown in Fig.~\ref{f:smfs}, comprises a single sheet, originating from
hybridized states of predominantly rare-earth / nickel $d$-character.
A closer look at the FS reveals that it indeed has strong potential for
nesting. In Fig.~\ref{f:smfs}b, a contour of the FS at several slices
through constant $c^*$ is shown, with the arrow indicating a nesting vector
of $(0.5, 0.52, 0)$. In order to further investigate the role of nesting,
the non-interacting susceptibility was investigated.  Neglecting matrix
elements, the non-interacting susceptibility, $\chi_{0}({\bf q},\omega)$
for wavevector ${\bf q}$ and frequency $\omega$ can be expressed as:
\begin{eqnarray}
\chi_{0}({\bf q},\omega) = \sum_{nn',{\bf k}} \frac{ f(\epsilon_{n,{\bf k}}) -
f(\epsilon_{n',{\bf k}+{\bf q}})}{\epsilon_{n,{\bf k}}-\epsilon_{n',{\bf k}+{\bf
q}} - \omega -i\delta },
\end{eqnarray}
where $f(\epsilon)$ is the Fermi function and the sum runs over bands $n$
and $n'$ through the Brillouin zone. The Brillouin zone integration is
performed using the tetrahedron method of Rath and Freeman \cite{rath75}.
As recently emphasized by Johannes
{\it et al.}, although the imaginary part of $\chi_{0}({\bf q})$
depends only on details of the electronic structure near the Fermi energy,
it is the {\it real} part of $\chi_{0}({\bf q})$ (which has contributions
coming from a bandwidth-sized window of energies) that could indicate a
CDW instability \cite{johannes08}. Both the real and imaginary parts of the
non-interacting susceptibility were calculated for SmNiC$_2$, and are shown
in Fig.~\ref{f:chi1}. It is noteworthy that there is a strong peak at the same
{\bf q} vector in both the real and imaginary parts of $\chi_{0}({\bf q})$.
Closer inspection reveals that the peak is at a wavevector of $(0.5,
0.56, 0)$, which is very close to the CDW vector ${\bf q}_1$ identified by
Shimomura {\it et al.}\ \cite{shimomura09}. It should be borne in mind when
comparing the experimental ${\bf q}_1$ and theoretical peak in $\chi_0({\bf
q})$ that our calculations have been performed at the experimental lattice
constant, rather than at the energy minimum of the LMTO calculation.
Nevertheless, excellent qualitative, and good quantitative
agreements are found between experiment and these band structure calculations.
Approximations beyond the
ASA (such as full-potential schemes) and a fully-relativistic treatment may
also improve the precise location of the peak in $\chi_{0}({\bf
q})$, and its agreement with the experimental ${\bf q}_1$. The $\chi_{0}({\bf q})$ was
also calculated along the $[1,1,1]$ direction (shown in Fig.~\ref{f:chi111})
and again a peak was found at a vector of $(0.5, 0.5, 0.5)$, which matches
the ${\bf q}_{\rm R}$ vector associated with the thermal diffuse scattering
by Shimomura {\it et al.}\ \cite{shimomura09}.

\begin{figure}[bt]
\includegraphics[width=1.00\linewidth,clip]{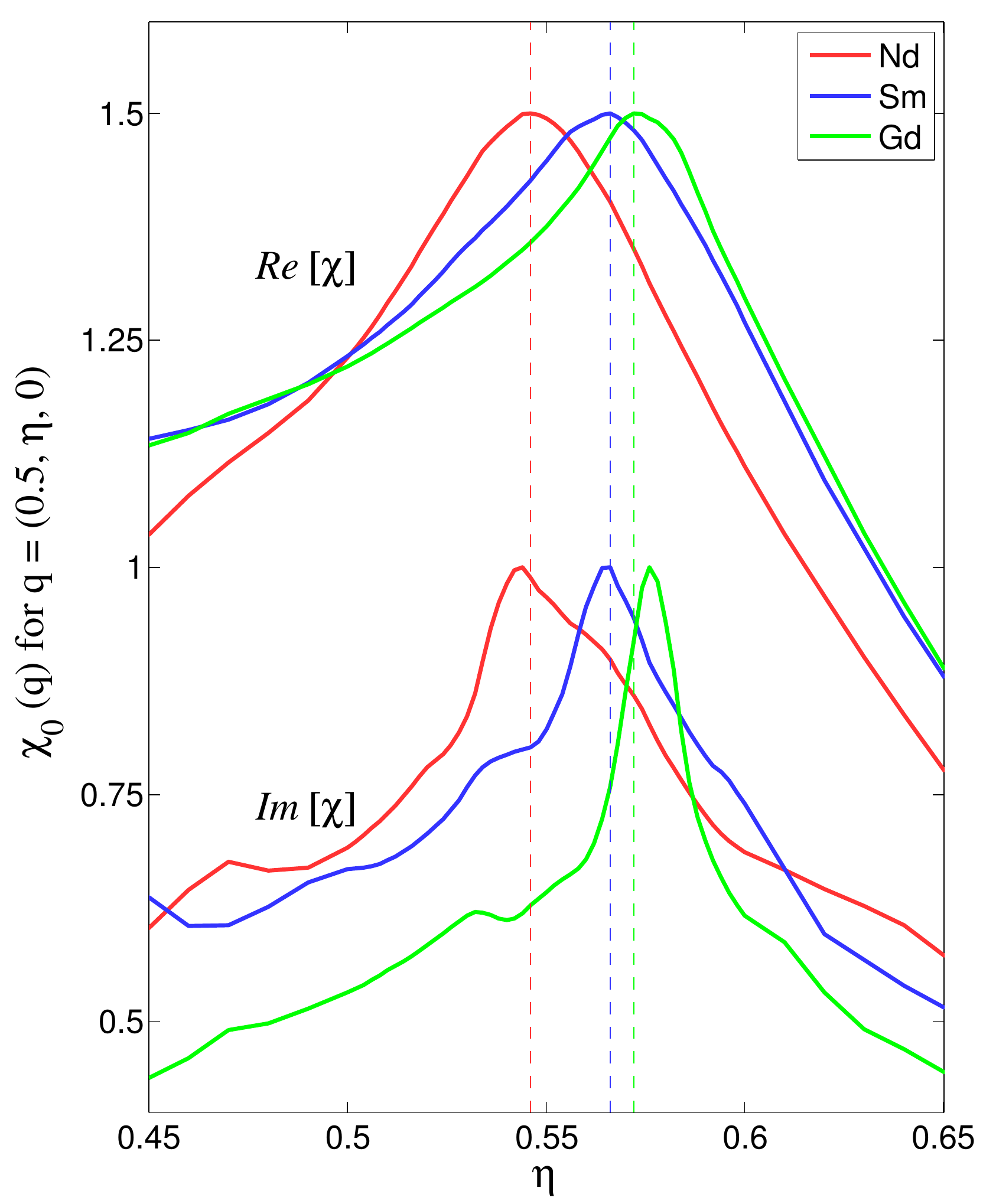}
\caption{\label{f:rplot} (color online) The real (top) and imaginary (bottom)
parts of $\chi_0({\bf q})$ for different rare-earth elements (Nd, Sm and
Gd), shown along the ${\bf q} = (0.5, \eta, 0)$ direction as a function of
$\eta$. The dashed vertical lines indicate the location of the peaks in the
real part of the susceptibility for each rare-earth.}
\end{figure}

Calculations for both NdNiC$_2$ and GdNiC$_2$ reveal similar electronic
structure and Fermi surfaces to the Sm compound, and the nesting properties
already identified for SmNiC$_2$ are still present in a visual inspection
of the FS. Furthermore, calculations of both the real and imaginary parts of
$\chi_0({\bf q})$ show comparable structure, the strong peak previously
identified for the Sm compound developing in ${\it Re}[\chi({\bf q})]$ for
${\bf q} = (0.5, 0.55, 0)$ and ${\bf q} = (0.5, 0.57, 0)$ for NdNiC$_2$
and GdNiC$_2$, respectively (see Fig.~\ref{f:rplot}).  Both the Nd and Gd
compounds exhibit anomalies in the resistivity that have been identified with
the emergence of a CDW \cite{murase04}. Although the wavevectors that define
these CDW states have not yet been investigated, these results suggest the
CDW structure is expected to be similar in nature to that already observed
for SmNiC$_2$, with a weak dependence on the particular rare-earth compound
(presumably associated with the lanthanide contraction).  The coincidence of
the peak structure in both the real and imaginary parts of the susceptilibity
for several different rare-earths, demonstrated in Fig.~\ref{f:rplot},
underlines the importance of the FS (through nesting) in these systems.

\begin{figure}[tb]
\includegraphics[width=1.00\linewidth,clip]{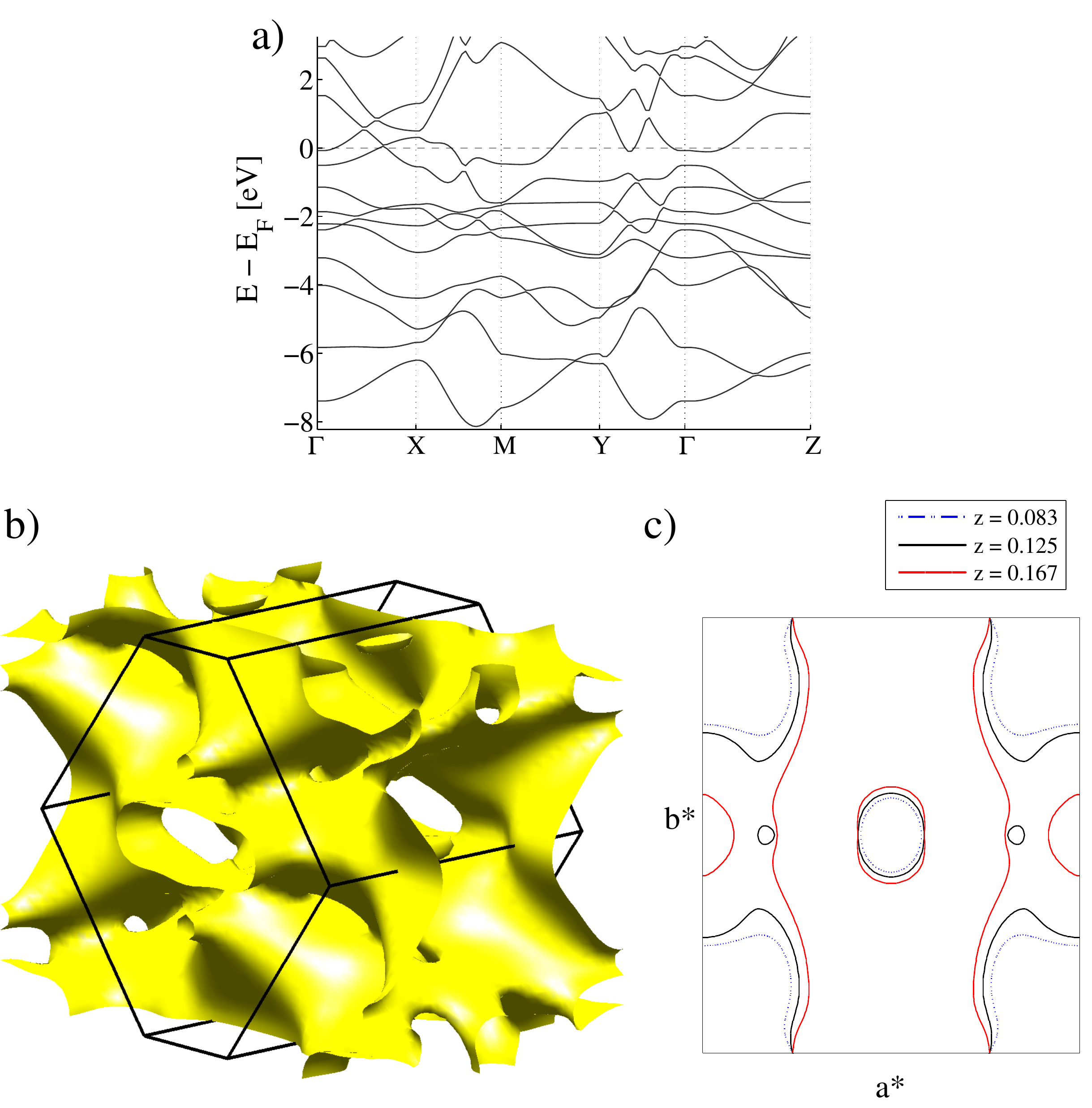}
\caption{\label{f:lafs} (color online) The electronic band structure (a)
and FS of LaNiC$_2$, (b) in 3D with the first Brillouin zone marked, and
(c) several contours as for Fig.\ \ref{f:smfs}.}
\end{figure}

\begin{figure}[tb]
\includegraphics[width=1.00\linewidth,clip]{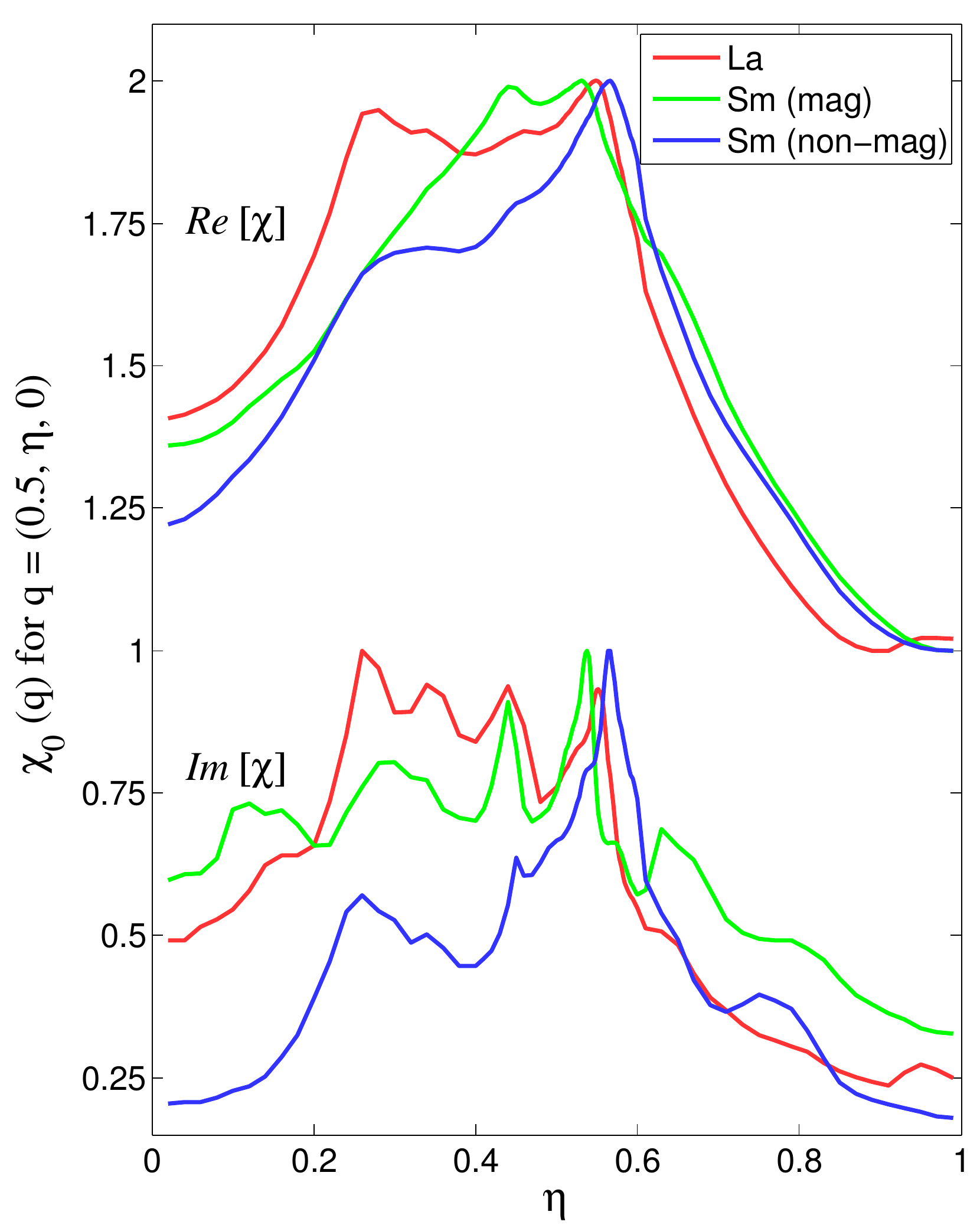}
\caption{\label{f:lamag} (color online) Real (top) and imaginary (bottom)
parts of $\chi_0({\bf q})$ for LaNiC$_2$ and ferromagnetic SmNiC$_2$ (at
the LMTO moment of $0.17\mu_{\rm B}$). For comparison, the $\chi_0({\bf
q})$ for non-magnetic SmNiC$_2$ shown in Fig.~\ref{f:chi1} is reproduced
here. Note the broad hump in $\chi_0({\bf q})$ for
both LaNiC$_2$ and ferromagnetic SmNiC$_2$, in contrast to the sharp peak
shown in Fig.~\ref{f:chi1} for non-magnetic SmNiC$_2$.}
\end{figure}

Conversely, LaNiC$_2$ shows no equivalent kink in its resistivity \cite{murase04}
for temperatures down to 12K, raising the question of whether this
compound shares the same FS topology, and more specifically, whether the
same nesting feature is present.  We have addressed this by calculating
the electronic structure and $\chi_0({\bf q})$ for LaNiC$_2$. The
density of states at the Fermi level is shown in Table~\ref{rareearthtable},
alongside comparative values for the other rare-earth compounds addressed in
this study. The FS, shown in Fig.~\ref{f:lafs}b, is found to be topologically
quite different, and the features of the FS that give rise to the nesting
in SmNiC$_2$ appear strongly distorted, as demonstrated by the constant
$c^*$ contours shown in Fig.~\ref{f:lafs}c. The ${\it Im} [\chi_0({\bf
q})]$ shows a broad series of ripples along the ${\bf q} = (0.5, \eta, 0)$
direction (see Fig.~\ref{f:lamag}), rather than the strong peak observed for
SmNiC$_2$, compounding the evidence of poor nesting. In the real part of the
susceptibility, also shown in Fig.~\ref{f:lamag}, the peak becomes spread out
over some $\sim 0.4 b^*$ of the Brillouin zone, which is unlikely to drive a CDW.
Calculations for SmNiC$_2$ in the LaNiC$_2$ structure yield similar results,
emphasizing that the transition of the FS from strong to weaker nesting is a
consequence of the structure, rather than the particular rare-earth substitution.

Spin-polarized calculations of the electronic structure have also been
performed for SmNiC$_2$ in order to investigate whether a similar transition in
the nesting qualities of the FS (as a consequence of the spin-splitting of the
FS) can explain the disappearance of the CDW state that accompanies the onset
of ferromagnetism. These calculations have been performed at the same lattice
constant used for non-magnetic SmNiC$_2$, since the structural transition
that has been reported at the Curie temperature \cite{murase04} is
small and does not break any additional symmetry.  Although a total ordered
moment of $0.32 \mu_{\rm B}$ is reported experimentally \cite{onodera98}, the
magnetic state of the localized Sm $f$ electrons has not yet been clarified.
Measurements of X-ray Magnetic Circular Dichroism on the $K$-edge of
Ni in SmNiC$_2$ at 5K have indicated that there is a finite moment on
the Ni site \cite{mizumaki06}, but a full analysis of the moment was not
possible. In our calculations, a {\em spin} moment of $0.17 \mu_{\rm B} /
{\rm Sm}$ develops, which is about half that reported experimentally, and
is predominantly located on the Ni site.  Inspection of the ferromagnetic FS
suggests that the different band filling in the majority sheet substantially
weakens the nesting, but it still remains to some extent in the minority FS.
Calculations of $\chi_0({\bf q})$ reveal a strong suppression of the $(0.5,
\eta, 0)$ peak in the imaginary part, whereas its real counterpart displays
a broad ridge over a large range of $\eta$, which would again be unlikely to
drive a CDW instability. Additional calculations at a fixed spin moment that
corresponds to the (total) experimental magnetic moment reveal this effect is
even stronger, and the broadening of the ridge structure in the real part of
$\chi({\bf q})$ is spread over almost half the Brillouin zone in the $(0.5,
\eta, 0)$ direction.

\section{Conclusion}
We have calculated the FS of several (La, Nd, Sm and Gd) members of the
$R$NiC$_2$ series, accompanied by calculations of both real and imaginary
parts of the non-interacting susceptibility $\chi_0({\bf q})$. For the
Nd, Sm and Gd members, peaks in the imaginary part of $\chi_0$, that can be
directly identified with corresponding nesting properties of the FS, still
persist in the real part of $\chi_0$, which is the relevant quantity for a
CDW instability. The wavevector $[0.5, \eta, 0]$ corresponding to these peaks
in $\chi_0({\bf q})$ is close ($\eta = 0.56$) to the experimental wavevector
($\eta = 0.52$) obtained from x-ray scattering measurements of SmNiC$_2$
\cite{shimomura09}. On the other hand, the FS of LaNiC$_2$ is found to be
markedly different, and the $\chi_0({\bf q})$ does not exhibit such strong
structure, providing a possible route towards explaining the apparent absence
of a CDW state for this particular compound. Spin-polarized calculations
for SmNiC$_2$ at two different magnetic moments ($0.17$ and $0.32 \mu_{\rm
B}$) demonstrate the gradual destruction of the nesting properties of the FS
with increasing moment. The strong peak in $\chi_0({\bf q})$ that is present
at zero moment becomes washed out as the spin-polarized bands separate, in
agreement with the suggestion of Shimomura {\it et al.}\ that the ferromagnetic
transition in this compound adversely affects the nesting properties of the
FS and leads to a destruction of the CDW state \cite{shimomura09}.

In summary, our findings suggest that for the Sm, Gd and
Nd compounds, the FS does indeed nest at the CDW wavevector recently
observed for SmNiC$_2$, and more importantly, this strong divergence
survives in the real part of the susceptibility. Furthermore, calculations of
two isostructural compounds for which CDW formation has not been observed
show substantially weaker divergences in their respective susceptibility,
lending support to the idea that FS nesting plays an important role in
deciding the fate of these systems.

\section*{Note added during revision}
During the preparation of this manuscript, a recent study of the electronic
structure of LaNiC$_2$ has been reported \cite{subedi09}, in which the authors
have relaxed the internal crystal structure. They go on to comment that the FS
they obtain is in excellent agreement with our results. In response to their
investigation, we have repeated the calculation of LaNiC$_2$ for these
parameters, and find no significant change in the FS.

\section*{Acknowledgements}
We would like to thank G.~Santi for his development of the susceptibility code,
and the EPSRC (UK) for financial support.


\begin{thebibliography}{100}
\bibitem{atpos}
The atomic positions used in the calculations were: $R$: (0, 0, 0.9936); Ni:
(0.5, 0, 0.608); C: (0.5, 0.155, 0.285); C: (0.5, -0.155, 0.285).
\end{thebibliography}
\end{document}